\documentclass[conference]{IEEEtran}
\usepackage{cite}
\usepackage{blindtext}
\usepackage{amsmath,amssymb,amsfonts}
\usepackage[ruled,linesnumbered]{algorithm2e}
\usepackage{graphicx}
\usepackage{textcomp}
\usepackage{algpseudocode}
\usepackage{xcolor}

\usepackage{booktabs}
\usepackage{xurl}
\usepackage{hyperref}
\usepackage{array}
\IEEEoverridecommandlockouts
\def\BibTeX{{\rm B\kern-.05em{\sc i\kern-.025em b}\kern-.08em
    T\kern-.1667em\lower.7ex\hbox{E}\kern-.125emX}}
\begin{document}
\title{Using Neural Networks for Novelty-based Test Selection to Accelerate Functional Coverage Closure}

\author{\IEEEauthorblockN{Xuan Zheng}
\IEEEauthorblockA{\textit{Trustworthy Systems Laboratory} \\
\textit{University of Bristol}\\
Bristol, UK \\
dq18619@bristol.ac.uk}
\and
\IEEEauthorblockN{Kerstin Eder}
\IEEEauthorblockA{\textit{Trustworthy Systems Laboratory} \\
\textit{University of Bristol}\\
Bristol, UK \\
kerstin.eder@bristol.ac.uk}
\and
\IEEEauthorblockN{Tim Blackmore}
\IEEEauthorblockA{\textit{50-60 Station Road} \\
\textit{SiFive}\\
Cambridge, UK \\
tim.blackmore@gmail.com}
}

\maketitle    
\begin{abstract}
Novel test selectors used in simulation-based verification have been shown to significantly accelerate coverage closure regardless of the number of coverage holes. This paper presents a configurable and highly-automated framework for novel test selection based on neural networks. Three configurations of this framework are tested with a commercial signal processing unit. All three convincingly outperform random test selection with the largest saving of simulation being 49.37\% to reach 99.5\% coverage. The computational expense of the configurations is negligible compared to the simulation reduction. We compare the experimental results and discuss important characteristics related to the performance of the configurations.
\end{abstract}

\begin{IEEEkeywords}
Simulation-Based Verification, Functional Coverage, Novelty Detection, Neural Network
\end{IEEEkeywords}

\maketitle

\section{Introduction}\label{s:introduction}
Simulation-based verification is a vital technique that is used to gain confidence in the functional correctness of digital designs. Traditionally, the quality of a generated test is measured by various coverage metrics obtained during simulation. 
Functional coverage records whether the specified functionality has been executed by simulated tests. In practice, functional coverage closure requires the generation and simulation of many tests.
The coverage gain that can be realised by random test simulation (the state of art technique used in the industry) sharply decreases along with verification progress. These barriers slow down the development of digital designs and become more severe because of the increasing complexity of modern digital designs.

In recent years, Machine Learning (ML) algorithms have gained ground in Coverage Directed Test Generation (CDG) to improve functional coverage~\cite{ioannides2012coverage}. The loop between coverage and test generation is empirically learned by ML algorithms to automatically alter the next tests to be run to target the coverage holes left by previous simulation rounds. A knowledge gap affecting the training phase is that for coverage holes, the simulated tests that can be used by ML algorithms as positive examples to learn do not exist. Therefore, ML algorithms cannot effectively bias test generation or filter out generated tests to target coverage holes in
the next simulation round. Rarely covered events are also affected in a similar way. This problem is more severe when verifying large-scale designs because the correlation between tests and coverage events is more complicated. To overcome this knowledge gap, syntax-based distance metrics have been used to identify coverage events that have been hit in syntactically close proximity to coverage holes~\cite{eder2007ilp}. However, syntactic proximity, e.g.\ based on the Hamming distance between coverage tuples, is no guarantee for semantic proximity. In~\cite{9474160}, an automatic scalable system based on the distance metrics to cluster coverage events is used to increase the hit times of rarely-hit and un-hit events. Nevertheless, the computational expense for the system to improve the coverage increases rapidly when the number of coverage holes rises.

On the other hand, novel test selectors have been shown to significantly accelerate coverage closure for simulation-based verification without being affected by the lack of positive examples\cite{bworld, guzey2008functional,chang2010online,chen2012novel }. This approach is based on the hypothesis that dissimilar tests (novel tests) are more likely to hit dissimilar functional coverage events. Therefore, simulating dissimilar tests is assumed to increase coverage more effectively than the simulation of randomly selected tests. The dissimilarity between tests can be captured by Novelty Detection (ND) techniques. The previous novel test selectors have the prominent merit that the training and prediction phases are independent of coverage space and only test information is required. In\cite{bworld}, an autoencoder is trained to reconstruct simulated tests and future tests with high reconstruction errors are deemed novel. Unlike the explicitly-designated distance metrics used in\cite{guzey2008functional,chang2010online,chen2012novel}, the reconstruction error is a more general way to represent how novel an input test is relative to the training set. Additionally, an explicit distance metric that can be generally used across the projects does not exist~\cite{guzey2008functional}.~\cite{9356340} demonstrates, in a Neural Network (NN), the hidden neurons for a novel input data have the values that have not been activated by the training set, which is also a general measurement for the novelty between tests. Another merit of using NN-based test selectors is that the training expense does not explode along with the increasing number of training samples in high-dimensional space.  

Inspired by the approaches of test selection and the prominent features of NN discussed above, this paper presents a Neural Network based Novel Test Selector (NNNTS) framework, of which the input is a test, and the output space can be configured to represent different definitions of novelty. NNNTS selects the tests generated from a Constraint Random Test Generator (CRTG) to ensure novel tests are prioritized during simulation.
Three configurations of the output space are given in the paper. The first configuration is to set NNNTS to reconstruct the input test in the output space, i.e. an autoencoder\cite{bworld}. The tests with high reconstruction errors are regarded as the most novel. The second configuration is to set NNNTS to predict the correlation between tests and coverage events. However, instead of biasing the test generator or filtering generated tests according to the prediction result, the novelty score calculated from the outputs of hidden neurons is assigned to each test~\cite{9356340}. We extend and improve~\cite{9356340} to cater for novel test selection.
The output space in the third configuration directly generates a non-linear score to indicate the novelty degree of un-simulated tests in the coverage space formed by simulated tests, which is originally proposed in our paper. The effectiveness of the configurations of NNNTS is demonstrated on an industrial-scale dataset. The experiment for each configuration of NNNTS is repeated multiple times with the randomly-chosen initialization parameters. This is to explore how the initialization parameters influence the performance and to avoid a coincidental result. Random test selection is also repeated in abundance to avoid coincidence. From the results, all the configurations can evidently accelerate the closure of functional coverage compared to random test selection though they perform differently from each other. We extract the important characteristics of NNNTS and discuss how they are related to the performance differences among the configurations. Although our innovative configuration (third configuration) outperformances the first configuration\cite{bworld} for reaching 99\% coverage level, the first configuration has a better performance for reaching 99.5\% coverage level. This suggests that different configurations may suit different coverage models though further researches need to be done. Nevertheless, the characteristics of NNNTS discussed in this paper may serve as an inspiration for others to generate a new configuration with a further improved performance compared to three configurations in this paper.

The merits of using NNNTS to select novel tests are:
\newline

1. The performance of NNNTS is not limited by the number of un-hit and rarely-hit coverage events.

2. Highly automated: The construction of NNNTS is independent of domain knowledge from a design under verification. The generated instances of test templates can be automatically transformed into test vectors driven to NNNTS. While the schemes for test vector encoding in\cite{guzey2008functional,chang2010online,chen2012novel} require many manual efforts.

3. Comparatively light computational expense: The computation expense of NNNTS can be negligible compared to the simulation reduction brought by it. Moreover, the training expense of previous test selectors\cite{guzey2008functional,chang2010online,chen2012novel} becomes dramatically expensive with the increasing number of training samples in high-dimensional space. This defect is alleviated in NNNTS.

4. Compared to other ND techniques\cite{pimentel2014review}, it can quantitatively compare how novel each un-simulated test is, relative to simulated tests instead of qualitatively predicting which un-simulated tests are novel. Furthermore, the novelty among tests is generally compared without an explicit distance metric.

5. NNNTS is easy to implement and maintain because it does not bias CRTG or modify generated tests.
\newline

The paper is structured as follows. Section~\ref{s:previous} reviews and compares the related work to unveil the research questions yet to solve and illustrate the inspirations for this paper. Section~\ref{s:Metho} illustrates how novel test selection is generally conducted, three important characteristics and the construction of NNNTS and the encoding scheme of test vectors. Section~\ref{s:Experimental} presents and discusses the experimental results of NNNTS in the particular simulation-based verification environment of a commercial design. The relation between the characteristics of NNNTS and the performance of the configurations is also expressed in this section. Section~\ref{s:Conclusion} concludes the paper and discusses future work.

\section{Previous Work}\label{s:previous}

Previous work has embedded ML algorithms in CDG to accelerate coverage closure using fewer tests than the random simulation~\cite{ioannides2012coverage}. During the training phase, ML empirically learns the correlation between the constraints used to generate simulated tests, simulated tests and the corresponding coverage data until the satisfying prediction performance of ML is achieved. Afterwards, the trained ML can be either used to bias the test generation or to filter generated tests with the aim to fill coverage holes or increase existing coverage.
However, for missing coverage events, the positive tests for the training do not exist and thus ML cannot effectively increase coverage as intended. This problem has been aware and researched in~\cite{eder2007ilp} and~\cite{9474160}. In~\cite{eder2007ilp}, Inductive Logic Programming (ILP) generates directives that target coverage holes by learning from related coverage clusters. The results show that the hit times of both rarely-hit and uncovered events can be increased though the design used in the experiment is fairly small. Besides, the syntax-based distance metric can limit the potential space of related events\cite{9218685}.~\cite{9474160} searches for the neighbour events close to rarely-hit and un-hit coverage events via the distance metrics. Then, a group of existing test templates that have the relevant parameters of hitting the neighbour events of the target event are selected and processed. A derivative-free optimization (DFO) technique is used to manipulate these test templates hoping to cover the target event. However, this framework requires extensive simulations when many uncovered events are involved in the process.
\setlength{\parskip}{0pt}

By contrast, novel test selectors have been shown to effectively accelerate coverage closure by arranging the simulation order of generated tests regardless of the number of coverage holes~\cite{guzey2008functional,chang2010online,chen2012novel,bworld}. It is assumed that the tests novel to those already simulated are more likely to execute functionality not yet covered and detect potential bugs. The responsibility of a novel test selector is to repeatedly select a batch of un-simulated tests for simulation, which are most novel to simulated tests. The training and execution of previous test selectors are de-coupled from coverage space, which makes them not suffer from the knowledge gap in Section~\ref{s:introduction}.

In~\cite{bworld}, an autoencoder is used to select novel tests. Each field in a test represents the value in each register that configures the functionality of the design. Autoencoder is trained to reproduce simulated tests and the novelty score is defined as the mean squared difference between the input and reproduced tests. A higher novelty score implies a higher portion of simulated tests dissimilar to the input test. The case study is based on the signal processing unit of commercial complexity with about 6,000 white-box coverage events. The coverage events are internal signals, which are hard to hit. Therefore, the case study demonstrates the scalability of using an autoencoder to select novel tests to accelerate the closure of functional coverage for a large-scale design. In particular, the test vector encoding is simple in~\cite{bworld}. Each feature in the vector is directly extracted from the value of a configuration register. For other test selectors based on One-Class Support Vector Machine
(OCSVM)~\cite{guzey2008functional,chang2010online,chen2012novel}, extra manual efforts and processes are needed to encode test vectors. Besides, a common kernel-induced distance metric does not exist in the OCSVM-based test selectors to capture the novelty between test vectors encoded by different schemes. While in~\cite{bworld}, the novelty is more generally represented by how different an input test is relative to the training set, instead of relying on a specific distance metric. In addition, the training expense of OCSVM-based test selectors increases dramatically with the increasing number of samples in high-dimensional space. This makes the repetitive training shown in Figure~\ref{fig:scheme} unacceptably costly when the number of simulated tests becomes high.

The authors of~\cite{9356340} associate the novelty in the input space with the output value of hidden neurons. Novel data samples are regarded to have outlier values for the neuron outputs. Inspired by~\cite{9356340}, we assume that the density in the input space can also be projected onto the output of hidden neurons and verify the assumption in \ref{s:Experimental}. In~\cite{gogri2020machine}, a supervised NN is also used to predict the correlation between tests and coverage. The result shows that the simulation of the tests with the low-confidence prediction can accelerate the increase of functional coverage compared to the random selection. Low-confidence prediction indicates the corresponding input to ML is quite different from the training set and thus it is a novel input \cite{mandelbaum2017distance}. However, for this confidence-based approach, it cannot quantitatively tell which un-simulated test is more novel to the training set. Therefore, this potential configuration is not extended in our framework.  

\section{Methodology}\label{s:Metho}

\subsection{Test Selection Loop}

\begin{figure}
    \centering
    \includegraphics[scale=0.30]{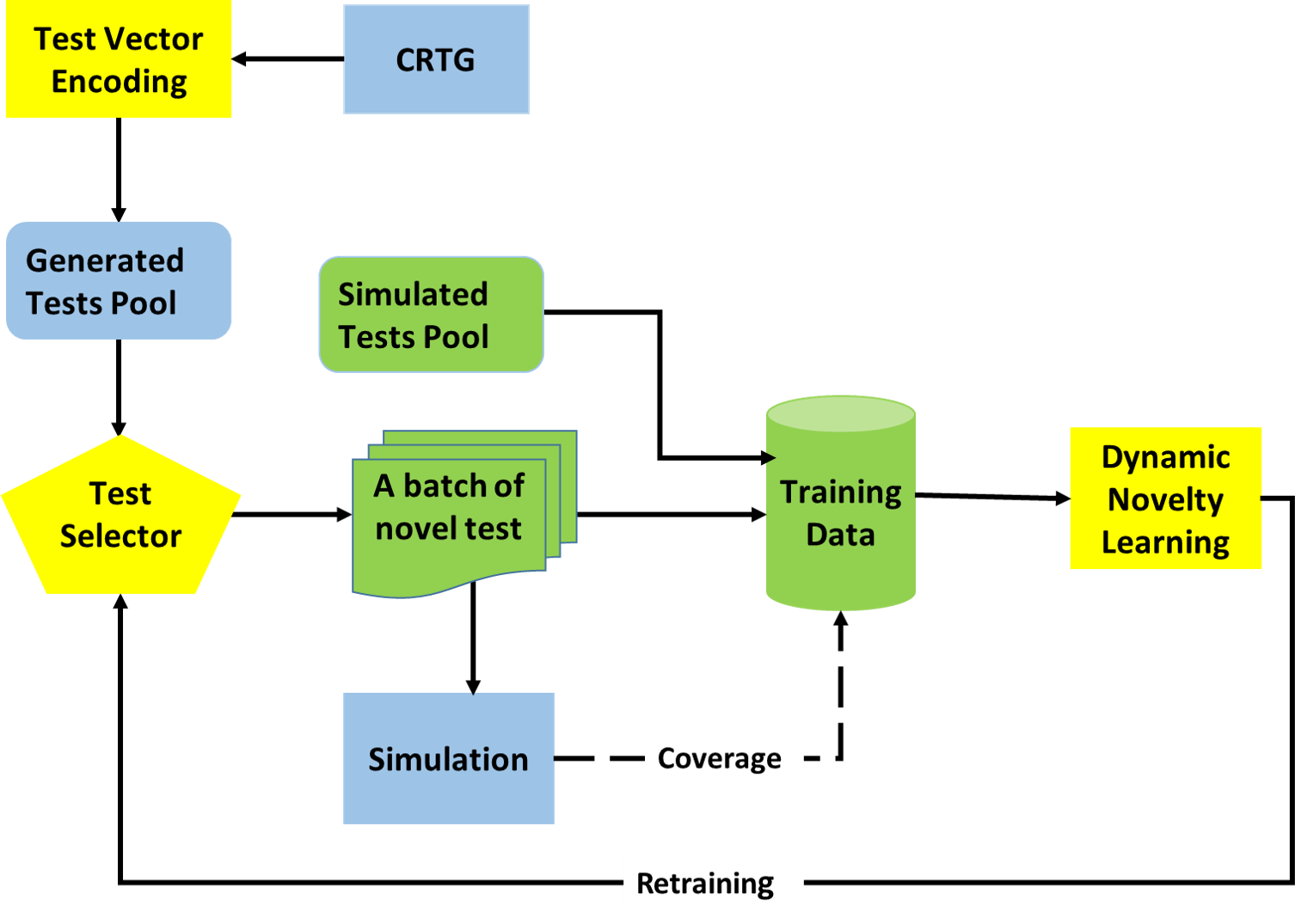}
    \caption{Novel Test Selection Loop}
    \label{fig:scheme}
\end{figure}

The principle of novel test selection is based on the hypothesis that dissimilar tests are more likely to exercise dissimilar functionality of the design. This can be evidenced by more coverage events being hit by a group of dissimilar tests than a group of similar tests~\cite{guzey2008functional}. Derived from the principle, a loop for continuously selecting novel tests for simulation-based verification is proposed in this section, as shown in Figure~\ref{fig:scheme}. 

First, the tests generated from CRTG are encoded as test vectors. A test vector is formed by various features, e.g. the features can be extracted from the values driven to configuration registers (configuring the functionalities of the design), sequences driven to the design, instructions executed on a processor and so on. These features already exist in the generated instances of test templates and can be automatically used to encode test vectors. Compared to the encoding schemes of test vector in ~\cite{guzey2008functional,chang2010online,chen2012novel} that require more manual intervention, our scheme is more concise and much less time-consuming. Encoded test vectors are then stored in Generated Test Pool. 

The test selector is initially trained with a small number of simulated tests before being embedded into the verification environment. In~\ref{s:Experimental}, the number of initial tests is set to 100. After the initial training, the pre-trained test selector is in the selection phase, in which the objective is to generate the novelty score for each test. Afterwards, a batch of the most novel tests relative to the simulated tests is selected. The batch size in~\ref{s:Experimental} is set to be 1000. Simulating a new batch of tests dynamically changes the definition of novel tests. To obtain the optimum performance of the test selector, retraining it with all the simulated tests after each simulation round is necessary before selecting a new batch of novel tests. The above process repeats until either a termination criterion is reached, or all the generated tests are simulated. The coverage information of simulated tests can also be logged in the simulated tests pool to form the training set if it is required by NNNTS.

\subsection{Construction of Test Selector}\label{s:Construction of Test Selector}

\begin{figure}
    \centering
    \includegraphics[scale=0.31]{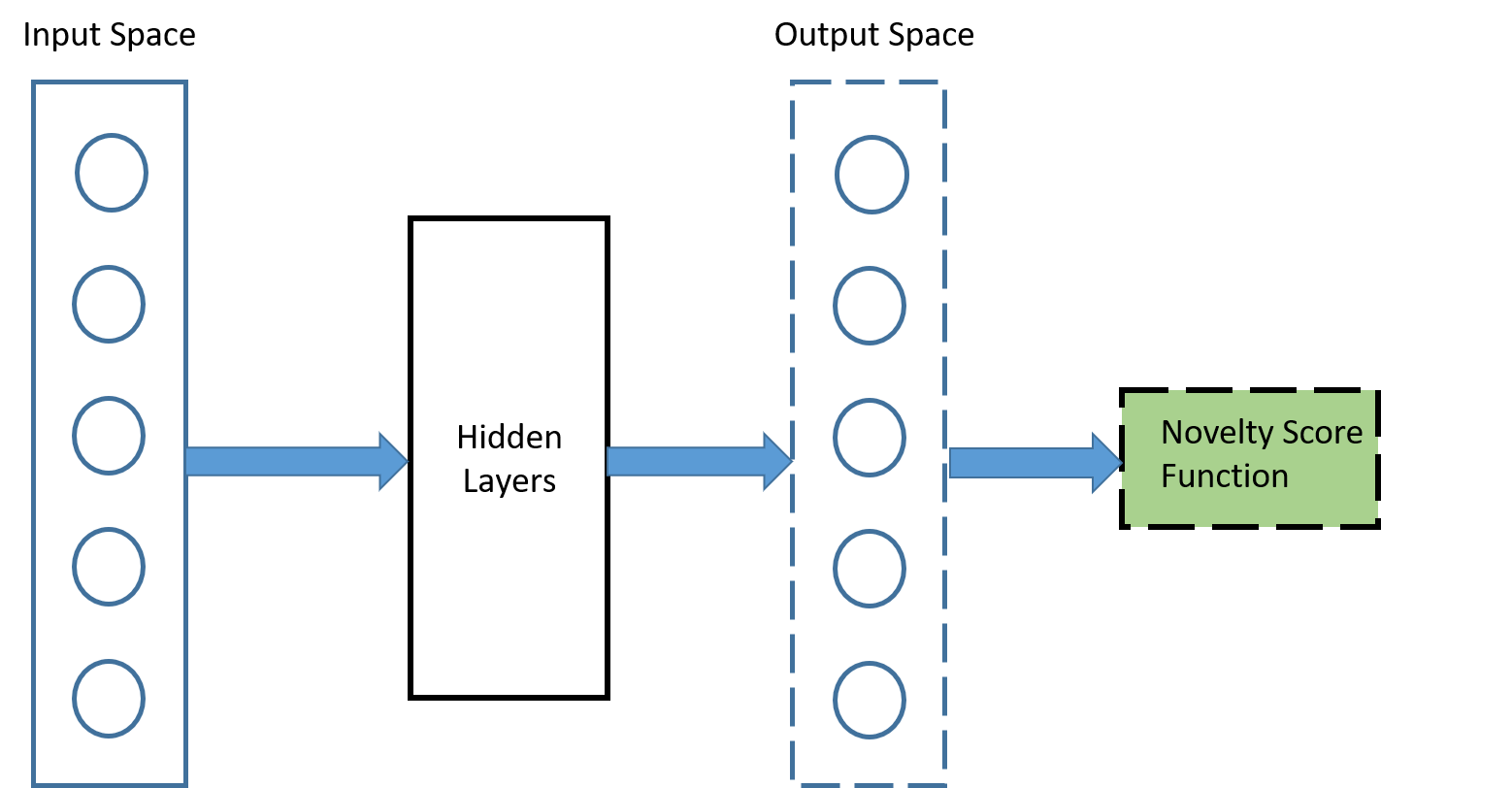}
    \caption{General Structure of NNNTS}
    \label{fig:structure}
\end{figure}

\subsubsection{General Structure of NNNTS}
The general structure of a configurable NNNTS is based on NN, as shown in Figure \ref{fig:structure}. A test vector is driven as the input to NNNTS and therefore the number of input space dimensions matches the number of features of a test vector. The neurons of adjacent layers are fully connected. The user-defined parameters for NNNTS include the number of neurons in each layer, the number of layers, and the activation functions, which can influence the performance of NN. However, fine tuning these parameters is not in the scope of this paper. We follow the basic guideline to configure the NNs, which is to keep the number of neurons in each hidden layer to be either twice, half of, or the same as the number in the previous layer. Instead, configuring the output space and the associated novelty functions is one of the main topics of this paper. 

We extract 3 characteristics of NNNTS as follows, identify what they are in each configuration in Section\ref{para:config} and discuss how they influence the performance of novel test selection by comparing the results of three configurations in Section \ref{s:result}:
\newline

1.	Novelty Detector: The smallest unit to conduct ND which is either a neuron in the output space or a hidden layer depending on the configuration of the output space.

2.  Novelty Score Function: Describes how novelty is measured in Novelty Detector and how novelty measured from each of them is gathered to output a final score to indicate how novel an input test is relative to simulated tests

3.	Simulation Information for Training: How the simulation information is used to train NNNTS
\newline

\subsubsection{Three configurations of NNNTS}\label{para:config}
Herein, Density NNNTS, Autoencoder NNNTS and Coverage-Novelty NNNTS respectively correspond to NNNTS with the output space being configured to three modes. For Density NNNTS, each neuron in the output space is configured to output the probability of a coverage event being hit by the input test, i.e.\ it is a logistic regression model that estimates the correlation between tests and coverage events. We assume that the novelty of an input test relative to the training set can be projected onto the output of each subsequent hidden neuron. Thus, Novelty Detectors in this configuration are all the hidden neurons. This idea is inspired by~\cite{9356340}, in which the distance between the input data and the training set is equal to the number of hidden neurons in the corner-activation areas. The corner-activation areas are the ranges of values that have never been activated by the training set and are reached by the input test in the execution phase. Thus, the corner-activation areas are low-density regions. Novelty Score Function for an un-simulated test in the input space is the sum of the differences between the values of hidden neurons corresponding to the input and their K-nearest neighbours (KNN) values. The search scope for KNN of a hidden neuron for an input test is within the values of that hidden neuron for simulated tests. We assume that a positive correlation exists between the summed differences from the output value of a hidden neuron to its KNN and the local sparsity of that neuron. This is formally illustrated in Algorithm~\ref{alg:DS}.  

Suppose we input an un-simulated test T to Density NNNTS and N is the novelty score for T. $O_{us}$ is the list of output values of all the hidden neurons for T. $O_{s}$ is the list of output values of all the hidden neurons for simulated tests (the training set). The index of $O_{s}$ is the index of hidden neurons. For instance, $O_{s}[0]$ is the value set that contains outputs of hidden neuron 0 for simulated tests. Similarly, $O_{us}[0]$ is the value of hidden neuron 0 for T. N\_K is the set of values in $O_{s}[i]$ which are KNN to $O_{us}[i]$  in Euclidean space. The sum of Euclidean distances from $O_{us}[i]$  to its KNN is the score component $n$ obtained from hidden neuron $i$. Compared to~\cite{9356340}, Density NNNTS provides a more detailed approach for calculating the novelty difference between the input samples rather than just recognizing the novel inputs. Moreover, the boundaries of corner-activation areas in \cite{9356340} are calculated from the training set. A training set containing outliers can cause the boundary values to become extreme and thus the corner-activation areas can be falsely expanded. An input outlier can be regarded as normal in this case. While Density NNNTS alleviates this problem by conducting the density estimation for the output values of each hidden neuron.

\begin{algorithm}
  \caption{Calculating a Novelty Score for an Un-simulated Test in Density NNNTS}\label{alg:DS}
    $N = 0$\;
    \For{\texttt{< $i$ in range(0,len($O_{s}$)>}}{
      \texttt{Find N\_K of $O_{us}[i]$ in $O_{s}[i]$}\;
      $n = 0$\;
      \For{\texttt{< $j$ in range(0,len($N\_K$)>}}{
        $n = n + D[O_{us}[i],N\_K[j]]$\;
      }
    $N = N + n$\;  
    }
\end{algorithm}

For the hidden layers in Density NNNTS, a Leaky ReLu function is adopted as the activation function rather than a ReLu function used in~\cite{9356340} for two reasons. First, a ReLu neuron may never be activated after a large negative gradient flows through the neuron, which causes the neuron to always output zero during the feedforward and backward propagation~\cite{xu2020reluplex}. Second, the outputs of a ReLu neuron for all the negative inputs are zero, which causes the loss of density information from the negative territory in the input space of the neuron. By contrast, the minimum value that a Leaky ReLu function can represent is negative infinity. Although Density NNNTS empirically learns the correlation between tests and coverage events, the test selection is not based on the prediction result of whether an input test can hit the coverage holes. The knowledge gap discussed in Section~\ref{s:introduction} is subtly circumvented by acquiring the novelty score of the input data in the output space of hidden neurons, transferring a prediction task to an ND task. 

On the other hand, NNNTS can also be configured to compress an input test into lower dimensions and then reconstruct the test from the compressed dimensions. Such a configuration of a NN is known as Autoencoder in deep learning areas and it is termed Autoencoder NNNTS in this paper. The output neuron for a reconstructed feature is used as a Novelty Detector. The mean squared difference that expresses the reconstruction error is regarded as Novelty Score Function as shown in Equation~\ref{eq:MSE}.

\begin{equation}
  \label{eq:MSE}
  \begin{aligned}
   N &= \frac{1}{n}\sum_{j=1}^{n}(I_j - O_j)^2
  \end{aligned}
\end{equation}

$I_j$ is the $j$th feature of an input test vector and $O_j$ is the jth reconstructed feature in the output space and $n$ is the total number of features. In the execution phase, the higher the reconstruction error is, the more novel the input test vector is. 

For Autoencoder and Density NNNTS, the coverage information of simulated tests is not directly utilized to select novel tests though in the latter approach it is used to train the NN. We propose an innovative configuration of the output space, in which a single neuron (Novelty Detector) directly estimates how novel an input test is in the coverage space constructed by the simulated tests. 
This is a supervised learning algorithm called Coverage Novelty NNNTS. We assume that for a simulated test, if a coverage event hit by the test is also frequently hit by other simulated tests, then the test is very similar to the other tests in that coverage-event dimension (an entry of $C_{array}$ in Algorithm 2). The overall dissimilarity of a simulated test in the coverage space is the sum of the dissimilarity in each coverage-event dimension. This is formally illustrated in Algorithm~\ref{alg:CN} (Novelty Score Function) where N is the coverage-novelty score for a simulated test and $n_{c_{i}}$ is the score component calculated from each coverage-event dimension.

\begin{algorithm}
    $N = 0$\;
    \For{\texttt{< $i$ in range(0,len($C_{array}$)>}}{
      \If{$C_{array}[i] = 0$}{$n_{c_{i}} = 0$\;}
      \Else{ $n_{c_{i}} = \frac{1}{{C_{HitTimes}[i]}*{\sqrt{C_{HitTimes}[i]}}} $\;            }
      $N = N + n_{c_{i}}$\;
    }
\caption{Generating a Training Label for a Simulated Test in Coverage-Novelty NNNTS}\label{alg:CN}
\end{algorithm}

In Algorithm~\ref{alg:CN}, $C_{array}$ is the vector that records whether each coverage event has been hit by the test requiring the training label and $C_{HitTimes}$ is the vector that records the times of each coverage event being hit by simulated tests. 
For example, $C_{array}[0] = 1$ represents the coverage event 0 is hit by the test and  $C_{HitTimes}[1] = 20$ means twenty simulated tests hit the coverage event 1. $n_{ci}$ is 0 if the test requiring the label does not hit coverage event $i$. N is the sum of $n_{ci}$ over all the coverage events. The denominator ${{C_{HitTimes}[i]}*{\sqrt{C_{HitTimes}[i]}}}$ in the else branch distributes more scores to the rarely-hit coverage events of a simulated test. 
Thus, the situation where the test that activates a few infrequently-hit coverage events is assigned a lower N than the test that activates many frequently-hit coverage events can be easily mitigated.
The labelling algorithm only includes the coverage information of simulated tests and thus does not require the test information for un-hit coverage events. The trained NN then is used to predict how novel an un-simulated test is relative to the coverage space of the simulated tests. The full correlation between all generated tests and the labels is only known after the simulation of all the tests. However, it is not the case that the correlation between tests and coverage is not known for coverage holes or extremely limited known for rarely-hit events. Section \ref{s:result}  confirms that Coverage-Novelty NNNTS can still select novel tests to accelerate the coverage increase before the correlation between the tests and labels is fully known.

\section{Experimental Evaluation}\label{s:Experimental}
\subsection{The experiment background}\label{s:Exp_background}
In this study, the design under verification is the Signal Processing Unit (SPU) of the Advanced Driver Assistance Subsystem (ADAS), which is the same design used in~\cite{bworld}. The SPU processes RADAR data received from sensors in the vehicle. Besides, the processing is highly configurable via a large number of configuration registers. Thus, the SPU is a complex, safety-critical and highly configurable IP.

The simulation-based method is used to verify the SPU. The simulation-based verification is metric-driven, constrained-random verification with the test generation and the simulation environment being decoupled. A generated test consists of a configuration of the SPU, stored in a database, and the RADAR data to be processed. Only the configuration is used as the features of a test vector, contributing to 290 raw features (290 configuration registers). Moreover, the features are automatically engineered, for example to condense large, data-centric configuration fields, leaving a total of 265 final features.

Afterwards, the engineered features are standardized with the library in Scikit-Learn~\cite{scikit-learn} before being presented to NNNTS. For the functional coverage model, there are 8409 white-box functional coverage events. This coverage model is different from the one used in~\cite{bworld}. The real-life project requires the 6-month simulation of nearly 2 million constrained random tests with almost 1,000 machines and EDA licenses. The simulation expense of each test is 2 hours on average. Moreover, several months of effort are paid by verification engineers to write constraints targeting coverage holes. For the experiment, 3076 tests that give 100\% functional coverage are saved from the completed project. In addition, 82335 tests are produced by a CRTG (i.e.\ as random as possible within the constraints of a valid test), giving a total of 85411 tests. These tests are shuffled thoroughly, encoded and then stored in the generated test pool in Figure~\ref{fig:scheme} before any selection and simulation of tests.

NNNTS of all the configurations are constructed by using Keras~\cite{chollet2015keras}. For each NNNTS in the experiment, a proportionate structure is configured. The configuration guideline is illustrated in~\ref{s:Construction of Test Selector}. The following vectors illustrate the number of layers and the number of neurons in each layer for the NNNTS: 
\newline

 Autoencoder-NNNTS: [265, 128, 64, 128, 265]
 
 Density-NNNTS: [265, 512, 265, 128, 50]
 
 Coverage-Novelty-NNNTS: [265, 265, 128, 64, 1]
\newline

Each element from the left of the vector to the right represents the number of neurons in each layer starting from the input layer. For instance, the Autoencoder-NNNTS vector represents a NN with five layers and the input and output layers respectively have 265 and 265 neurons. The first hidden layer has 128 neurons, with another two hidden layers following it. The output space of Density-NNNTS consists of the neurons only corresponding to the 50 randomly-selected coverage events. By reducing the number of neurons in the output space, the number of neurons in the hidden layers can correspondingly be reduced. Thus, the expense of calculating the novelty score for Density NNNTS can also be reduced. However, the option of 50 randomly-selected coverage events is not the result of fine-tuning for the optimum performance of the NNNTS. The N\_K in Algorithm 1 is set to 15 and the rationale is also to reduce the training expense.

\subsection{Experimental Results}\label{s:result}
In this experiment, 100 tests are randomly sampled from all the generated tests to initialize NNNTS before any round of test selection. The number of tests to be selected in each round is set to be 1000. Retraining occurs after the simulation of the selected tests. All the simulated tests and the associated coverage information (if necessary) are used to form the retraining set.
To avoid a coincidental result, the experiment for each NNNTS is repeated 10 times with different initial states and the random selection is repeated 5000 times. The initial states are different in terms of initial tests (the number of initial tests is fixed to 100) and initial weights of NNs. Nevertheless, the same initial tests are used to initialize all the configurations in one repetition round of the experiment. Specifically, for the experiment of Density NNNTS, 50 coverage events are randomly re-sampled for each repetition of the experiment. The coverage for 10-time simulations of initial tests is between 56.30\% and 44.15\%.

\begin{figure}
    \centering
    \includegraphics[scale=0.53]{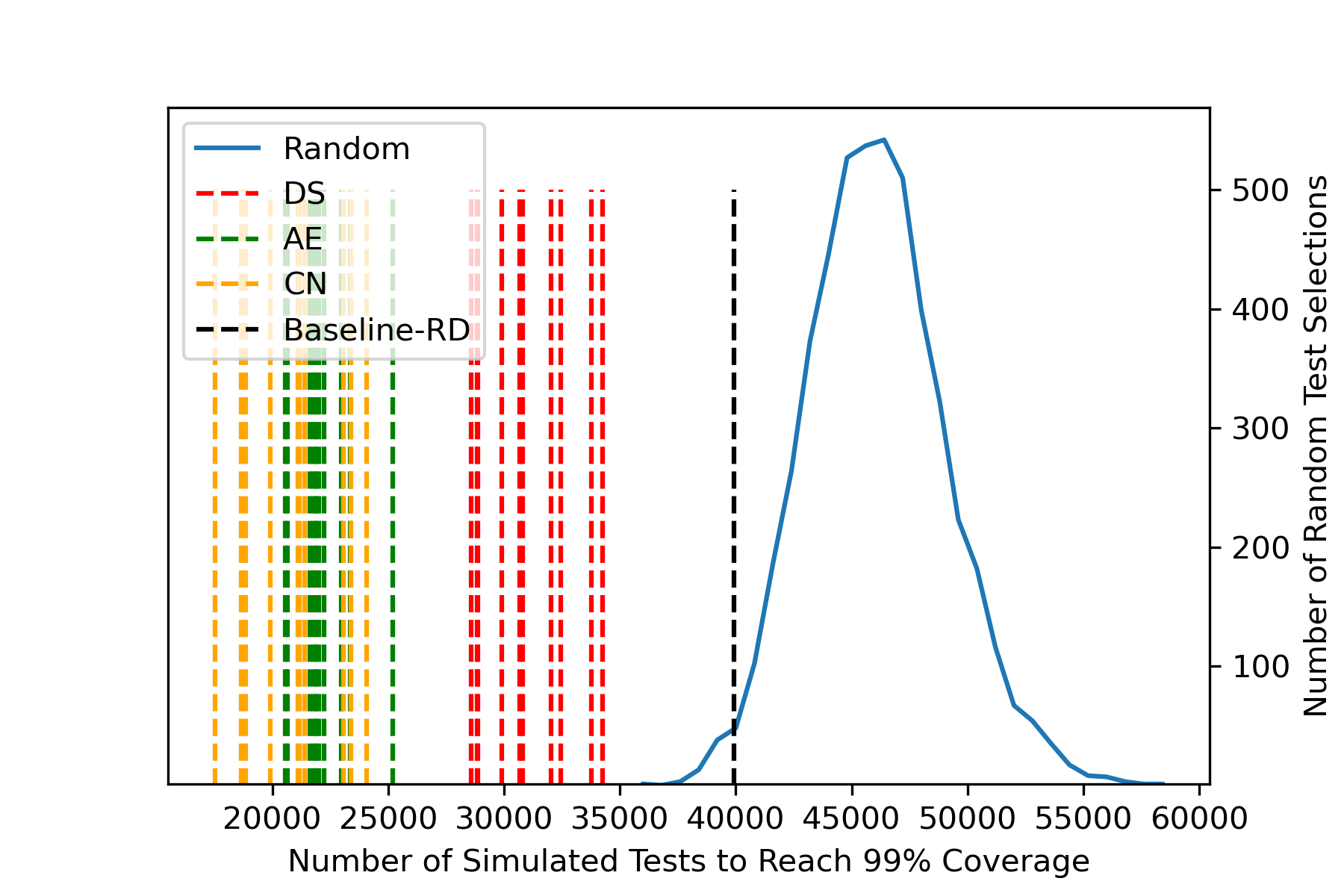}
    \caption{Result of Using NNNTS to improve Verification Efficiency (99\%)}
    \label{fig:plot_99}
\end{figure}

\begin{figure}
    \centering
    \includegraphics[scale=0.53]{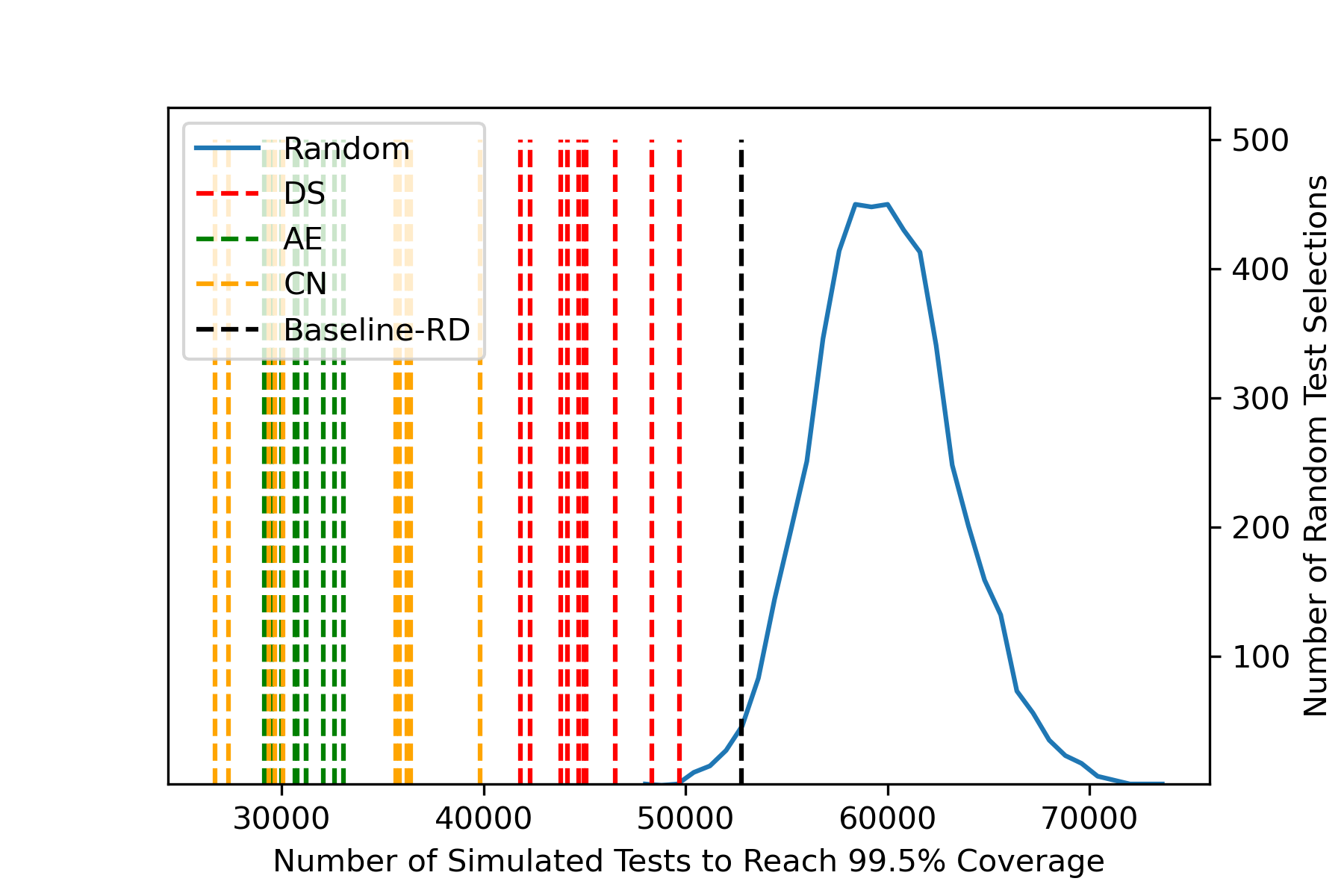}
    \caption{Result of Using NNNTS to improve Verification Efficiency (99.5\%)}
    \label{fig:plot_99_5}
\end{figure}

From the result of the experiment, all the configurations of NNNTS can continuously select more efficient tests than the random selection during test simulation. However, the number of required tests to achieve higher coverage levels is more concerning. In the random selection, the number of redundant tests being simulated aggressively increases with the increase of coverage. Figures \ref{fig:plot_99} and \ref{fig:plot_99_5} show the performance of random test selection (Random) compared to the simulations of tests selected by Density (DS), Coverage-Novelty (CN) and Autoencoder (AE) NNNTS for two high coverage levels. The horizontal axis represents the number of required tests to achieve the intent coverage level. The vertical axis is only read for the random-selection group, which represents the number of random selections that need the corresponding horizontal-axis number of tests to reach the coverage level. For instance, more than 60 random selections require about 40,000 tests to reach 99\% coverage. In each of the figures, the 50th-best random selection to reach the intended coverage is chosen as the baseline (Baseline-RD) to be compared with the simulations accelerated by NNNTS. This depicts the benefits of deploying NNNTS compared to 99\% of the 5000 random selections. 
In terms of achieving coverage levels 99\% and 99.5\%, all the configurations of NNNTS can noticeably accelerate the increase of functional coverage compared to the random selection. The average computational expenses brought by Coverage Novelty, Autoencoder and Density NNNTS to achieve 100\% coverage are respectively 19 hours, 21 hours and 33 hours when running the experiment on an Intel Xeon Gold 6248R CPU. Considering the average simulation time of a test being 2 hours in this experiment, the additional expense brought by deploying NNNTS is negligible compared to the simulation reduction as shown in Figures \ref{fig:plot_99} and \ref{fig:plot_99_5}.

Table\ref{table:NNNTS} gives a more detailed performance comparison between three NNNTS and random selection. The most and least columns respectively represent the most and least savings out of 10 runs of each NNNTS relative to the 50th-best random selection to reach the intended coverage level. The average saving for each NNNTS is also presented. Coefficient of Variation (CV) is used as an indicator of performance stability, which is defined as the ratio of the standard deviation to the mean. The lower CV is, the more stable the performance is. 

Autoencoder NNNTS has a great number of Novelty Detectors of which each captures the novelty in one feature of an input test vector. The simulated tests with all the features are used to train the NN. Besides, the mean squared difference is a mature Novelty Score Function that unifies the novelty obtained from each Novelty Detector. These characteristics contribute to the most stable performances of Autoencoder NNNTS for reaching 99\% and 99.5\% coverage. For Coverage-Novelty NNNTS, not only all the simulated tests are included in the training set, but also their corresponding coverage information is merged in the Novelty Score Function. The ND directly conducted in the coverage space leads to Coverage-Novelty NNNTS having the largest values for the largest, least and average savings for reaching 99\% coverage. However, the performance of Coverage-Novelty NNNTS becomes relatively less stable for reaching 99.5\% coverage level. We assume this is caused by the NN only having one Novelty Detector in the output space for such a large coverage space consisting of more than 8,400 events. To improve the performance of Coverage-Novelty NNNTS, the coverage events could be clustered via the distance metrics and ML clustering techniques \cite{9218685} first and then a Novelty Detector can be allocated to each of the event clusters. This can also improve the controllability to increase the hit times for a specific group of similar events. For Density NNNTS, although all the hidden neurons are used as Novelty Detectors, simply summing the novelty obtained from each detector can be a naive way of representing the final novelty score for a test. Moreover, increasing the number of coverage events to construct the output neurons could improve the performance because more simulation information is obtained in the training. However, the computational expense also increases and thus the cost-effectiveness of doing so should be evaluated.

In this experiment, 99.5\% coverage is the point where the correlation between tests and the remaining coverage holes becomes extremely complicated. At this point, directed tests crafted by experienced verification engineers are also simulated along the simulation accelerated by NNNTS to target the remaining coverage holes (around 80 coverage events for our example).    

\begin{table}[!t]
\caption{Result of Using NNNTS to improve Verification Efficiency} 
\centering 
\begin{tabular}{|l|c|c|c|c|} 
\hline\hline 
Savings & 99\% (Most) & 99\% (Least) & 99\% (Average) & CV \\ [0.05ex] 
\hline 
CN & 56.97\% & 39.70\% & 47.62\% & 10.89\% \\ 
AE & 48.48\% & 36.91\% & 44.33\% & 7.26\% \\
DS & 28.45\% & 14.23\% & 22.33\% & 21.84\% \\ [1ex] 
\hline 
\noindent
Savings & 99.5\% (Most) & 99.5\% (Least) & 99.5\% (Average) & CV  \\ [0.05ex] 
\hline 
CN & 49.37\% & 24.48\% & 38.01\% & 21.55\% \\ 
AE & 44.78\% & 37.34\% & 41.19\% & 5.62\%  \\
DS & 20.72\% & 5.80\% & 14.44\% & 30.57\% \\ [1ex] 
\hline 
\end{tabular}
\label{table:NNNTS} 
\end{table}

\section{Conclusion}\label{s:Conclusion}
This paper proposes a novel test selector based on NN to achieve faster closure of functional coverage than the random test selection in a simulation-based verification environment. The performance of the test selector is not limited by the number of rarely-hit and un-hit coverage events. Besides, it is highly automated, comparatively computationally cheap, easy to maintain and provides a general method to quantitatively compare the novelty between tests. We have considered three configurations of the test selector. The first configuration is based on Autoencoder while the second configuration utilizes the novelty obtained from the values of hidden neurons to reflect the novelty in the input space. The third configuration calculates the novelty scores for the input tests in the coverage space formed by the simulated tests. However, the configuration of the output space is not limited to the three modes illustrated in this paper. In the experiment, we demonstrate that the test selectors can select the tests bringing more cumulative coverage than the random selection in the verification environment for the commercial SPU. The performance comparison between the configurations reveals how the three extracted characteristics of NNNTS influence the process of test selection.

The input features of NNNTS in the experiment are derived from the static configuration of the SPU. For a future design, including the transactions dynamically driven to the design in the input feature space is our next step. By doing so, we hope that NNNTS can capture the temporal logic between the transactions.
\medskip

\section*{Acknowledgment}

This research work forms a part of a PhD project funded by Infineon, Bristol. We would like to express our sincere gratitude to James Buckingham from Infineon, Bristol, for his invaluable contributions as a proofreader and for providing valuable feedback on this paper. We would also like to extend our deep thanks to Infineon, Bristol for kindly providing the SPU and associated verification data, enabling us to evaluate the performance of our novel test selectors.

\bibliographystyle{abbrv}
\bibliography{bibliography}

\begin{thebibliography}{10}

\bibitem{chang2010online}
P.-H. Chang, D.~Drmanac, and L.-C. Wang.
\newblock Online selection of effective functional test programs based on
  novelty detection.
\newblock In {\em 2010 IEEE/ACM International Conference on Computer-Aided
  Design (ICCAD)}, pages 762--769. IEEE, 2010.

\bibitem{chen2012novel}
W.~Chen, N.~Sumikawa, L.-C. Wang, J.~Bhadra, X.~Feng, and M.~S. Abadir.
\newblock Novel test detection to improve simulation efficiency—a commercial
  experiment.
\newblock In {\em 2012 IEEE/ACM International Conference on Computer-Aided
  Design (ICCAD)}, pages 101--108. IEEE, 2012.

\bibitem{chollet2015keras}
F.~Chollet et~al.
\newblock Keras.
\newblock \url{https://keras.io}, 2015.

\bibitem{eder2007ilp}
K.~Eder, P.~Flach, and H.-W. Hsueh.
\newblock Towards automating simulation-based design verification using ilp.
\newblock In S.~Muggleton, R.~Otero, and A.~Tamaddoni-Nezhad, editors, {\em
  Inductive Logic Programming}, pages 154--168, Berlin, Heidelberg, 2007.
  Springer Berlin Heidelberg.

\bibitem{9474160}
R.~Gal, E.~Haber, W.~Ibraheem, B.~Irwin, Z.~Nevo, and A.~Ziv.
\newblock Automatic scalable system for the coverage-directed generation (cdg)
  problem.
\newblock In {\em 2021 Design, Automation \& Test in Europe Conference \&
  Exhibition (DATE)}, pages 206--211, 2021.

\bibitem{9218685}
R.~Gal, H.~Kermany, A.~Ivrii, Z.~Nevo, and A.~Ziv.
\newblock Late breaking results: Friends - finding related interesting events
  via neighbor detection.
\newblock In {\em 2020 57th ACM/IEEE Design Automation Conference (DAC)}, pages
  1--2, 2020.

\bibitem{gogri2020machine}
S.~Gogri, J.~Hu, A.~Tyagi, M.~Quinn, S.~Ramachandran, F.~Batool, and
  A.~Jagadeesh.
\newblock Machine learning-guided stimulus generation for functional
  verification.
\newblock In {\em Proceedings of the Design and Verification Conference
  (DVCON-USA), Virtual Conference}, pages 2--5, 2020.

\bibitem{guzey2008functional}
O.~Guzey, L.-C. Wang, J.~Levitt, and H.~Foster.
\newblock Functional test selection based on unsupervised support vector
  analysis.
\newblock In {\em 2008 45th ACM/IEEE Design Automation Conference}, pages
  262--267. IEEE, 2008.

\bibitem{9356340}
D.~Hond, H.~Asgari, and D.~Jeffery.
\newblock Verifying artificial neural network classifier performance using
  dataset dissimilarity measures.
\newblock In {\em 2020 19th IEEE International Conference on Machine Learning
  and Applications (ICMLA)}, pages 115--121, 2020.

\bibitem{ioannides2012coverage}
C.~Ioannides and K.~I. Eder.
\newblock Coverage-directed test generation automated by machine learning--a
  review.
\newblock {\em ACM Transactions on Design Automation of Electronic Systems
  (TODAES)}, 17(1):1--21, 2012.

\bibitem{mandelbaum2017distance}
A.~Mandelbaum and D.~Weinshall.
\newblock Distance-based confidence score for neural network classifiers.
\newblock {\em arXiv preprint arXiv:1709.09844}, 2017.

\bibitem{scikit-learn}
F.~Pedregosa, G.~Varoquaux, A.~Gramfort, V.~Michel, B.~Thirion, O.~Grisel,
  M.~Blondel, P.~Prettenhofer, R.~Weiss, V.~Dubourg, J.~Vanderplas, A.~Passos,
  D.~Cournapeau, M.~Brucher, M.~Perrot, and E.~Duchesnay.
\newblock Scikit-learn: Machine learning in {P}ython.
\newblock {\em Journal of Machine Learning Research}, 12:2825--2830, 2011.

\bibitem{pimentel2014review}
M.~A. Pimentel, D.~A. Clifton, L.~Clifton, and L.~Tarassenko.
\newblock A review of novelty detection.
\newblock {\em Signal processing}, 99:215--249, 2014.

\bibitem{bworld}
R.~H. Tim~Blackmore and S.~Schaal.
\newblock {Novelty-driven verification: Using machine learning to identify
  novel stimuli and close coverage}.
\newblock
  \url{https://www.dropbox.com/s/iulpk8kba3f7xxn/DVCon%20US%202021_Proceedings-FINAL.zip?dl=0&file_subpath=%2FPapers%2F7060.pdf},
  2021.
\newblock Proceedings of the 2021 Design and Verification Conference (Virtual).

\bibitem{xu2020reluplex}
J.~Xu, Z.~Li, B.~Du, M.~Zhang, and J.~Liu.
\newblock Reluplex made more practical: Leaky relu.
\newblock In {\em 2020 IEEE Symposium on Computers and Communications (ISCC)},
  pages 1--7. IEEE, 2020.

\end{thebibliography}

\end{document}